\documentclass[twocolumn]{IEEEtran} 
\usepackage{latexsym}                                 
\usepackage[dvips]{graphicx}                          
\usepackage{amsmath}                                  
\usepackage{amssymb}
\usepackage{subfigure}

\newcommand{\erf}{\mathop{\mathrm{erf}}}
\newtheorem{theorem}{Theorem}
\newtheorem{lemma}{Lemma}

\newtheorem{corollary}{Corollary}

\begin{document}
\bibliographystyle{IEEEbib}
\title{Reconciliation of a \\Quantum-Distributed Gaussian Key\thanks{This work has been submitted to the IEEE for possible publication. Copyright may be transferred without notice, after which this version may be superseded.}}

\author{Gilles Van Assche\thanks{G. Van Assche is with the Ecole Polytechnique, CP 165, Universit\'e Libre de Bruxelles, 1050 Brussels, Belgium (e-mail: gvanassc@ulb.ac.be).}, Jean Cardinal\thanks{J. Cardinal is with the Facult\'e des Sciences, CP 212, Universit\'e Libre de Bruxelles, 1050 Brussels, Belgium (e-mail: jcardin@ulb.ac.be).} and Nicolas J. Cerf\thanks{N. Cerf is with the Ecole Polytechnique. He is also with the Information and Computing Technologies Research Section, Jet Propulsion Laboratory, California Institute of Technology, Pasadena, CA 91109 (e-mail: ncerf@ulb.ac.be).}}

\maketitle

\begin{abstract}
Two parties, Alice and Bob, wish to distill a binary secret key out of a list of correlated variables that they share after running a quantum key distribution protocol based on continuous-spectrum quantum carriers. We present a novel construction that allows the legitimate parties to get equal bit strings out of correlated variables by using a classical channel, with as few leaked information as possible. This opens the way to securely correcting non-binary key elements. In particular, the construction is refined to the case of Gaussian variables as it applies directly to recent continuous-variable protocols for quantum key distribution.
\end{abstract}


%
\section{Introduction}
%

With the advent of quantum key distribution (QKD), sometimes also called quantum cryptography, it is possible for two remote parties, Alice and Bob, to securely agree on secret information that shall later be used as a key for encrypting messages \cite{benn84,benn92:prl,benn92:sciam,gisi02:qc}. Although most QKD schemes make use of a discrete modulation of quantum states, such as BB84 \cite{benn84}, some recent protocols \cite{cerf00:qdgk,gros02:coherent} use a 
continuous modulation of quantum states, thus producing continuous random variables. In particular, in \cite{gros02:nature}, a QKD scheme based on the Gaussian modulation of quantum coherent states is demonstrated, which generates correlated Gaussian variables at Alice's and Bob's sides. The construction of a common secret key from discrete variables partly known to an adversary has been a long studied problem \cite{benn88:pa,bras93,maur93,benn95:pa}. However, in order to bring the intrinsically continuous QKD experiments up to getting a usable secret key, such key construction techniques needed to be adapted to Gaussian variables.

In QKD, the quantum channel that Alice and Bob use to create a secret key is not deemed to be perfect. Noise will necessarily make Alice's and Bob's values different. Furthermore, the laws of quantum mechanics imply that eavesdropping also causes extra discrepancies, making the eavesdropper detectable. To overcome this, one can correct errors by using some \emph{reconciliation} protocol, carried out over a public authenticated channel \cite{bras93,maur93}. Yet, this does not entirely solve the problem as an eavesdropper can gain some information about the key while Alice and Bob exchange their public reconciliation messages. Fortunately, such gained information can then be wiped out, at the cost of a reduction in the secret key length, using another protocol called \emph{privacy amplification} \cite{benn88:pa,benn95:pa}.

Current reconciliation and privacy amplification protocols are aimed at correcting and distilling strings of bits. However, the recently developed continuous-variable QKD schemes cannot be complemented efficiently with such discrete protocols. This paper proposes an extention of these protocols in the case of non-binary -- and in particular Gaussian -- key elements.  

%
\section{Quantum Distribution of a Gaussian Key}
%

In QKD, Alice and Bob use a quantum channel in order to share secret random data (a secret key) that can then be used for exchanging encrypted information. Since its inception, QKD has traditionally been developed with discrete quantum carriers, especially quantum bits (implemented e.g., as the polarization state of single photons). Yet, it has been shown recently that the use of continuous quantum carriers is advantageous in some situations, namely because high secret key bit rates can be attained~\cite{gros02:coherent}. The postprocessing of the raw data produced by such continuous-variable protocols therefore deserves further investigation.

As we shall see, the security of QKD fundamentally relies on the fact that the measurement of \emph{incompatible} variables inevitably affects the state of a quantum system. In a scheme such as BB84, Alice sends random key elements (e.g., key bits) to Bob using either one of two conjugate sets of quantum information carriers. Alice randomly chooses one of the two sets of carriers, encodes a random key element using this set, and sends it to Bob. On his side, Bob measures the received quantum state assuming either set was used at random. The two sets of quantum information carriers are designed in such a way that measuring the wrong set yields random uncorrelated results (i.e., the two sets are conjugate). Therefore, Bob will measure correctly only half of the key elements Alice sent him, not knowing which ones are wrong. After the process, Alice reveals which set of carriers she chose for each key element, and Bob is then able to discard all the wrong measurements, the remaining data making the key.

An eavesdropper (Eve) can of course intercept the quantum carriers and try to measure them. However, like Bob, Eve does not know in advance which set of carriers Alice chose for each key element. A measurement will yield irrelevant results about half of the time, and thereby disturb the state of the carrier. Not knowing if she has a correct value, Eve can decide to retransmit or not a quantum carrier with the key element she obtained. Discarding a key element is useless for Eve since this sample will not be used by Alice and Bob to make the key. However, if she does retransmit the state (even though it is wrong half of the time), Alice and Bob will detect her presence by an unusually high error rate between their key elements. QKD works because Bob has the advantage, over Eve, of being able to talk to Alice over a classical authenticated channel in order to select a common key and discard Eve's partial knowledge on it.

The continuous-variable QKD protocols described in \cite{cerf00:qdgk,gros02:coherent} take advantage of a pair of canonically conjugate continuous variables such as the two quadratures $X_1$ and $X_2$ of the amplitude of a mode of the electromagnetic field, which behave just like position $x$ and momentum $p$ \cite{scully}. The uncertainty relation $\Delta X_1\, \Delta X_2 \ge 1/4$ then states that it is impossible to measure with full accuracy \emph{both} quadratures of a single mode, $X_1$ and $X_2$. This can be exploited by associating the two sets of quantum information carriers with $X_1$ and $X_2$, respectively. For example, in the protocol \cite{cerf00:qdgk}, these two sets of carriers essentially behave like 2D Gaussian distributions in the $(X_1,X_2)$ plane. In set 1, the carriers are shaped as $N(x,\sigma_1) \times N(0,1/4\sigma_1)$, with $\sigma_1<1/4$ corresponding to the squeezing of $X_1$ \cite{scully}. Here, $x$ is the key element Alice wishes to send, and is itself distributed as a Gaussian: $x \sim N(0,\Sigma_1)$. In set 2, the carriers are similar but $X_1$ and $X_2$ are interchanged, that is, $N(0,1/4\sigma_2) \times N(x,\sigma_2)$, with $\sigma_2<1/4$. The raw key information is thus encoded sometimes in $X_1$ and sometimes in $X_2$, and the protocol resembles a continuous version of BB84. In contrast, in \cite{gros02:coherent}, two Gaussian raw key elements $x_1$ and $x_2$ are simultaneously encoded in a coherent state shaped as $N(x_1, 1/2) \times N(x_2, 1/2)$ in the $(X_1,X_2)$ plane. Bob can however only measure one of them, not both, so that only one Gaussian value $x=x_{1 \ \text{or} \ 2}$ is really transmitted. Eve, not knowing which one Bob will measure, necessarily disturbs $x_1$ when attempting to infer $x_2$ and vice-versa, and she in general disturbs both to some extent whatever the trade-off between acquired knowledge and induced disturbance she chooses. 

In all these continuous-variable protocols, the vacuum noise fluctuations of the transmitted states are such that Bob's measurement will not give him the exact value $x$ chosen by Alice, even in absence of eavesdropping and with a perfect measurement apparatus. The noise is Gaussian and additive, allowing us to model the transmission as a Gaussian channel. The amplitude of the noise can be estimated by Alice and Bob when they compare a subset of their exchanged values. Any noise level beyond the intrinsic fluctuations must be attributed to Eve, giving an estimate on the amount of information $I(X;E)$ that she was able to infer in the worst case \cite{cerf00:qdgk,gros02:coherent,gros02:nature}. This information, along with the information Eve gains by monitoring the reconciliation protocol, must then be eliminated via privacy amplification.

Finally, note that Alice must strictly respect $x \sim N(0,\Sigma_{1 \ \text{or} \ 2})$ or $(x_1, x_2) \sim N(x_1, 1/2) \times N(x_2, 1/2)$. She may not choose a codebook $x(k)$ from some discrete alphabet to $\mathbb{R}$ that displays the same variance. The resulting distribution would not be Gaussian, and Eve would be able to take advantage of this situation. For example in \cite{cerf00:qdgk}, measuring the correct or the wrong set must yield statistically indistinguishable results. If not the case, Eve would be able to infer whether she measured the correct set of carriers and adapt her strategy to this knowledge.

%
\section{Problem Description}
%

\subsection{Problem Statement}

The two parties each have access to a distinct random variable, namely $X$ for Alice and $X'$ for Bob, with non-zero mutual information $I(X;X')>0$. This models the quantum modulation and measurement of a QKD scheme, but other sources of common randomness could as well be used. When running the same QKD protocol several times, the instances of $X$ (resp. $X'$) are denoted $X_1\dots X_l$ (resp. $X'_1\dots X'_l$) for the time slots $1\dots l$, and are assumed independent for different time slots. The outcomes are denoted with the corresponding lower-case letters. An eavesdropper Eve also has access to a random variable $E$, resulting from tapping the quantum channel. These are also considered independent for different time slots, hence assuming individual attacks \cite{gisi02:qc}.

The goal of the legitimate parties is to distill a secret key, i.e., to end up with a shared binary string that is unknown to Eve. We assume as a convention that Alice's outcomes of $X$ will determine the shared key $K(X)$. It is of course not a problem if the roles are reversed, as required in \cite{gros02:nature}. The function $K(X)$ is chosen to be discrete, even if $X$ is continuous in nature, and this aspect is discussed below. 

In principle, secret key distillation does not require separate reconciliation and privacy amplification procedures, but it is much easier to use such a two-step approach.

First, reconciliation consists in exchanging reconciliation messages over the public authenticated classical channel, collectively denoted $C$, so that Bob can recover $K(X_{1\dots l})$ from $C$ and $X_{1\dots l}$. By compressing $K(X_{1\dots l})$, Alice and Bob can obtain about $lH(K(X))$ common random bits.

Then, privacy amplification can be achieved by universal hashing \cite{cart79,benn95:pa}. Starting from $K(X_{1\dots l})$, the decrease in key length is roughly equal to $lI(K(X);E)+|C|$, as shown in \cite{benn95:pa,cach97,maur00:strong}, where $|C|$ is the number of bits exchanged and where $I(K(X);E)$ is determined from the disturbance measured during the QKD procedure. Privacy amplification therefore does not need special adaptations in our case, as the existing protocols can readily be used.

Maximizing the net secret key rate $H(K(X))-I(K(X);E)-l^{-1}|C|$ involves to take all possible eavesdropping strategies into account during the optimization, which is very difficult in general. Instead, we notice that $I(K(X);E) \leq I(X;E)$, the latter being independent of the reconciliation procedure. Hence, we wish to devise a procedure that produces a large number of fully secret equal bits, hence to maximize $H(K(X))-l^{-1}|C|$.

\subsection{Discrete vs Continuous Variables}
\label{sec:dvsc}

It is shown in \cite{cerf00:qdgk,gros02:coherent,gros02:nature} that working with continuous quantum states as carriers of information naturally leads to expressing information in a continuous form. It is therefore natural to devise an all-continuous cryptographic processing. Nevertheless, we found more advantageous to distill a discrete secret key than a continuous one, and these aspects are now discussed.

First, a continuous secret key would need to be used along with a continuous version of the one-time pad, which is possible \cite{shan43}, but would most certainly suffer from incompatibilities or inefficiencies with regard to current technologies and applications. Furthermore, it is much more convenient to rely on the equality of Alice's and Bob's values in the discrete case, rather than dealing with bounded errors on real numbers. The resulting secret key is thus chosen to be discrete.

Second, the reconciliation messages can either be continuous or discrete. Unless the public authenticated classical channel has infinite capacity, exchanged reconciliation messages are either discrete or noisy continuous values. The latter case introduces additional uncertainties into the protocol, which quite goes against our purposes. Furthermore, a noisy continuous reconciliation message would less efficiently benefit from the authentication feature of the reconciliation channel. Hence, discrete reconciliation messages are preferred.

Third, the choice of a discrete final key also induces discrete effects in the protocols, which makes natural the choice of a continuous-to-discrete conversion during reconciliation. Call $x$ the original Gaussian value that Alice sent, $x'$ the Gaussian value as received by Bob, and $k$ the resulting discrete key element. The process of reconciliation and privacy amplification can be summarized as functions $k=f_A(x, c)$ and $k=f_B(x', c)$, where $c$ indicate the exchanged messages. As both $k$ and $c$ are to be taken in some finite set, these two functions define each a finite family of subsets of values that give the same result: $S_{k c} = \{ x \ : \ f_A(x, c)=k \}$ and $S'_{k c} = \{ x' \ : \ f_B(x', c)=k \}$. The identification of the subset in which $x$ (or $x'$) lies is the only data of interest -- and can be expressed using discrete variables -- whereas the value within that subset does not affect the result and can merely be considered as noise.

Finally, the discrete conversion does not put a fundamental limit on the resulting efficiency. It is possible (see Sec.~\ref{sec:sec}) to bring $|C|$ as close as desired to $lH(K(X)|X')$, giving almost $I(K(X);X')$ secret bits per raw key element. Also, one can define $K(X)$ as a fine-grained quantizer so that $I(K(X);X')$ can be made arbitrarily close to $I(X;X')$ \cite{cover}. On the other hand, no continuous protocol can expect Alice and Bob to share more secret information than what they initially share $I(X;X')$.

For all the reasons stated above, our reconciliation protocol mainly consists of exchanging discrete information between the two communicating parties so that they can deduce the same discrete representation from the real values they initially share.

%
\section{Sliced Error Correction}
%
\label{sec:sec}

Sliced error correction (SEC) is a generic reconciliation protocol that corrects strings of non-binary elements. It gives, with high probability, two communicating parties, Alice and Bob, equal binary digits from a list of correlated values. Just like other error correction protocols, it makes use of a public authenticated channel. The underlying idea is to convert Alice's and Bob's values into strings of bits, apply a bitwise correction protocol (BCP) as a primitive and take advantage of all available information to minimize the number of exchanged reconciliation messages.

The key feature of this generic protocol is that it enables Alice and Bob to correct errors that are not modeled using a binary symmetric channel (BSC), although using a BCP that is optimized for a BSC.

To remain general, Alice and Bob can process multi-dimensional key values and group them into $d$-dimensional vectors. In the sequel, $X$ and $X'$ denote $d$-dimensional variables, taking values in what is defined as the raw key space, i.e., ${\mathbb{R}}^d$ for Gaussian variables. When explicitly needed by the discussion, the dimension of the variables is noted with a $\cdot^{(d)}$ superscript.

To define the protocol, we must first define the slice functions. A slice $S(x)$ is a function from Alice's raw key space to $GF(2)$. A vector of slices $S_{1\dots m}(x)=(S_1(x), \dots, S_m(x))$ is chosen so as to map Alice's raw key elements to a discrete alphabet of size at most $2^m$. A vector of slices will convert Alice's raw key elements into binary digits, that is, $K(X)=S_{1\dots m}(x)$. 

Each of the slice estimators $\tilde{S}_1(x')$, $\tilde{S}_2(x', S_1(x))$ \dots $\tilde{S}_m(x', S_1(x), \dots, S_{m-1}(x))$ defines a mapping from Bob's raw key space and from Alice's slices of lower indexes to $GF(2)$. These will be used by Bob to guess $S_i(X)$ the best he can given his knowledge of $X'$ and of the slice bits previously corrected.

The construction of the slices $S_i(X)$ and their estimators depends on the nature and distribution of the raw key elements. These aspects are covered in a following section, where we apply the SEC to our Gaussian key elements.

Let us now describe our generic protocol, which assume that Alice and Bob defined and agreed on the functions $S_i$ and $\tilde{S}_i$.
\begin{itemize}
\item From her $l$ key elements $x_1 \dots x_l$, Alice prepares $m$ strings of bits using the defined slices $(S_1(x_1), \dots, S_1(x_l))$, \dots, $(S_m(x_1), \dots, S_m(x_l))$. She starts with the first one: $(S_1(x_1), \dots, S_1(x_l))$.
\item Bob constructs a string of bits from $x'_1 \dots x'_l$ using his slice estimator $\tilde{S}_1$: $(\tilde{S}_1(x'_1), \dots, \tilde{S}_1(x'_l))$.
\item Alice and Bob make use of a chosen BCP so that Bob aligns his bit string on Alice's.
\item For each subsequent slice $i$, $2 \leq i \leq m$, Alice takes her string $(S_i(x_1), \dots, S_i(x_l))$, while Bob constructs a new string using his slice estimator $\tilde{S}_i$ applied to his values $x'_1 \dots x'_l$ and taking into account the correct bit values of the previous slices $S_1(x_1), \dots, S_2(x_1), \dots, S_{i-1}(x_l)$. Again, Bob aligns his bit string to Alice's using the chosen BCP.
\item For Alice, the resulting bitstring is simply the concatenation of the $m$ $l$-bit strings: $S_{1\dots m}(x_{1 \dots l})$. For Bob, the shared bitstring is the same as Alice's, obtained from the previous steps.
\end{itemize}

The goal of SEC is to correct errors by disclosing as few information as possible on the key shared by Alice and Bob. However, one does not expect a protocol running with strings of finite length and using finite computing resources to achieve the Shannon bound $I(X;X')$ exactly. Yet, it is easy to show that SEC is indeed asymptotically efficient, that is, it reaches the Shannon bound in terms of leaked information when the number of dimensions $d$ (i.e., the input alphabet size) goes to infinity.

A famous theorem by Slepian and Wolf \cite{slep73} shows the achievability rate regions for encoding correlated sources. In the context of SEC, this means that, with $d$ sufficiently large, there exist slice functions such that disclosing the first $r=\lfloor dH(K(X^{(1)})|{X'}^{(1)})+1\rfloor$ slices $S_{1\dots r}(X^{(d)})$ is enough for Bob to recover the $m-r$ remaining ones and reconstruct $S_{1\dots m}(X^{(d)})$ with arbitrarily low probability of error. An alternate proof is proposed in Appendix~\ref{app:asympt}.

It is necessary here to quantize $X'$, as Slepian and Wolf's theorem assumes discrete variables. As shown in \cite{cover}, $X'$ can be approximated as accurately as necessary by a discrete variable $\hat{X}'$, with $H(K(X)|\hat{X}') \to H(K(X)|X')$.

%
\section{Analysis of Sliced Error Correction}
%

Let us now analyze the amount of information leaked on the public channel during SEC. Clearly, this will depend on the primitive BCP chosen. This aspect will be detailed in a following section.

If not using SEC, one can in theory use encoding of correlated information \cite{slep73} to achieve, when $l \to \infty$,
\begin{equation}
\label{eq:i0}
l^{-1}|C| = I_0 \triangleq H(S_{1\dots m}(X)|X').
\end{equation}
When using slices, however, the BCP blindly processes the bits calculated by Alice $S_i(X)$ on one side and the bits calculated by Bob $\tilde{S}_i(X', S_{1\dots i-1}(X))$ on the other. The $l$ bits produced by the slices are of course independent from time slot to time slot. Assuming a perfect BCP,
\begin{equation}
\label{eq:is}
l^{-1}|C| = I_s \triangleq \sum_{i=1}^{m} H(S_i(X)|\tilde{S}_i(X', S_{1\dots i-1}(X))) \geq I_0.
\end{equation}
The inequality follows from the fact that $H(S_{1\dots m}(X)|X')=\sum_{i}H(S_i(X)|X',S_{1\dots i-1}(X))$ and that the term in the sum cannot decrease if replaced by $H(S_i(X)|\tilde{S}_i(X', S_{1\dots i-1}(X)))$.
The primitive BCP can be optimized to work on a binary symmetric channel (BSC-BCP), thus blindly assuming that the bits produced by the slices and the slice estimators are balanced. Assuming a perfect BSC-BCP,
\begin{equation}
\label{eq:ie}
l^{-1}|C| = I_e \triangleq \sum_{i=1}^{m} h(e_i) \geq I_s,
\end{equation}
with $h(e) = -e\log e - (1-e)\log(1-e)$ and $e_i=\Pr[S_i(X) \neq \tilde{S}_i(X', S_{1\dots i-1}(X))]$. The inequality follows from Fano's inequality \cite{cover} applied to a binary alphabet. In practice, a BSC-BCP is expected to disclose a number of bits that is approximately proportional to $h(e)$, i.e., $(1+\xi)h(e)$ for some overhead constant $\xi$. An explicit construction of slice estimators applying the expression of $I_e$ in Eq.~(\ref{eq:ie}) is examined next.

\subsection{Maximum Likelihood Slice Estimators}

The error probability in slice $i$ can then be expressed as the probability that Bob's slice estimator yields a result different from Alice's slice:
\begin{gather}
e_i = P_{S_i \tilde{S}_i}^{01} + P_{S_i \tilde{S}_i}^{10}\text{, with} \\
P_{S_j \dots S_i \tilde{S}_i}^{\beta_j \dots \beta_i b} = \int_{{\mathcal{D}}_{S_j \dots S_i \tilde{S}_i}^{\beta_j \dots \beta_i b}} p(x, x') dx dx'\text{, and} \\
\begin{split}
{\mathcal{D}}_{S_j \dots S_i \tilde{S}_i}^{\beta_j \dots \beta_i b} = \{ (x, x') \ &: \ S_j(x)=\beta_j \land \dots \land S_i(x)=\beta_i \\ &\land \ \tilde{S}_i(x', S_{1\dots i-1}(x))=b \}.
\end{split}
\end{gather}

Maximizing the global efficiency of the slice estimators is not a simple task because the efficiency of a slice estimator $\tilde{S}_i$ recursively depends on all previous estimators $\tilde{S}_{j<i}$. For this reason, our goal here is simply to minimize each $e_i$, of which $h(e_i)$ is an increasing function for $0 \leq e_i < \frac{1}{2}$, by acting only on $\tilde{S}_i$. This results in an explicit expression for $\tilde{S}_i(x', S_1(x), \dots, S_{i-1}(x))$, see Eq.~(\ref{eq:mle}).

An individual probability $P_{S_i \tilde{S}_i}^{ab}$ can be expanded as a sum of smaller probabilities over all possible values $\beta_{j<i}$ of the previous slices, namely
\begin{equation}
\label{eq:PsmallPs}
P_{S_i \tilde{S}_i}^{ab} = \sum_{\beta_{1 \dots i-1}} P_{S_1 S_2 \dots S_{i-1} S_i \tilde{S}_i}^{\beta_1 \beta_2 \dots \beta_{i-1} a b}.
\end{equation}

Each of these terms can be further expanded as
\begin{gather}
\label{eq:Pbbbbab}
P_{S_1 \dots S_i \tilde{S}_i}^{\beta_1 \dots a b} = \int_{{\mathcal{B}}_{S_1 \dots S_{i-1} \tilde{S}_i}^{\beta_1 \dots \beta_{i-1} b}} P_{S_1 \dots S_i}^{\beta_1 \dots a}(x') dx'\text{, with} \\
P_{S_1 \dots S_{i-1} S_i}^{\beta_1 \dots \beta_{i-1} a}(x') = \int_{{\mathcal{A}}_{S_1 \dots S_{i-1} S_i}^{\beta_1 \dots \beta_{i-1} a}} p(x, x') dx\text{,} \\
{\mathcal{A}}_{S_i\dots S_j}^{\beta_i\dots \beta_j} = \{ x : S_i(x)=\beta_i \land \dots \land S_j(x)=\beta_j \}\text{, and} \\
{\mathcal{B}}_{S_1 \dots S_{i-1} \tilde{S}_i}^{\beta_1 \dots \beta_{i-1} b} = \{ x' \ : \ \tilde{S}_i(x', \beta_{i\dots i-1})=b \}.
\end{gather}

From this, it is easy to show that a slice estimator $\tilde{S}_i$ minimizes $e_i$ if it has the form
\begin{equation}
\label{eq:mle}
\tilde{S}_i(x', \beta_{1\dots i-1}) =
\left\{ \begin{array}{ll}
0 & \textrm{if } P_{S_{1\dots} S_i}^{\beta_{1\dots} 0}(x') > P_{S_{1\dots} S_i}^{\beta_{1\dots} 1}(x'), \\
1 & \textrm{otherwise,}
\end{array} \right.
\end{equation}
except for cases where the probabilities are equal or over some zero-measure set. To minimize $e_i=P_{S_i \tilde{S}_i}^{01}+P_{S_i \tilde{S}_i}^{10}$, one can thus take advantage of the independence of smaller terms in (\ref{eq:PsmallPs}) and minimize them individually. From Eq.~(\ref{eq:Pbbbbab}), the terms $P_{S_1 \dots S_i \tilde{S}_i}^{\beta_1 \dots a a}$, for a correct guess, and $P_{S_1 \dots S_i \tilde{S}_i}^{\beta_1 \dots a \bar{a}}$, for a wrong guess, result from the integration of the same function over two different sets, namely ${\mathcal{B}}_{S_1 \dots S_{i-1} \tilde{S}_i}^{\beta_1 \dots \beta_{i-1} a}$ and ${\mathcal{B}}_{S_1 \dots S_{i-1} \tilde{S}_i}^{\beta_1 \dots \beta_{i-1} \bar{a}}$. Therefore, the domain of correct guesses should simply cover all subsets in which the integrand is larger, and leave the smaller parts to the domain of wrong guesses. Eq.~(\ref{eq:mle}) is simply the maximum likelihood principle, expressed for slice estimators.

Note that when using Eq.~(\ref{eq:mle}), the bit error rate $e_i$ can be evaluated as
\begin{equation}
e_i = \sum_{\beta_{1\dots i-1}} \int \min \left( P_{S_{1\dots} S_i}^{\beta_{1\dots} 0}(x'), P_{S_{1\dots} S_i}^{\beta_{1\dots} 1}(x') \right) dx'.
\end{equation}

\subsection{Bitwise Correction Protocols}

To be able to use sliced error correction, it is necessary to chose a suitable BCP. There are first two trivial protocols that are worth noting. The first one consists in disclosing the slice entirely, while the second does not disclose anything. These are at least of theoretical interest with the asymptotical optimality of SEC: It is sufficient for Alice to transmit entirely the first $r=\lfloor dH(K(X^{(1)})|{X'}^{(1)})+1\rfloor$ slices and not transmit the remaining $m-r$ ones.

A BCP can consist in sending syndromes of error-correcting codes, see e.g., \cite{prad99:discus}. In binary QKD protocols, however, an interactive reconciliation protocol is often used, such as Cascade \cite{bras93,chen01,sugi00,yama00} or Winnow \cite{butt02}. In practice, interactivity offers overwhelmingly small probability of errors at the end of the protocol, which is valuable for producing a usable secret key.

Let us briefly analyze the cost of Cascade, which consists in exchanging parities of various subsets of bits \cite{bras93}. Let $A, B \in GF(2)^{l}$ be respectively Alice's and Bob's binary string of size $l$ constructed from some slice $S_i$ and its estimator $\tilde{S}_i$. After running Cascade, Alice (resp. Bob) disclosed $RA$ (resp. $RB$) for some matrix $R$ of size $n \times l$. They thus communicated the parities calculated over identical subsets of bit positions. The matrix $R$ and the number $n$ of disclosed parities are not known beforehand but are the result of the interactive protocol and of the number and positions of the diverging parities encountered. The expected value of $n$ is $n \approx l(1+\xi)h(e)$, where $e = \Pr[A_j \neq B_j]$ is the bit error rate, and $\xi$ is some small overhead factor.

If $A$ and $B$ are balanced and are connected by a BSC, the parities $RA$
give Eve $n$ bits of information on $A$, but $RB$ does not give any extra information since it is merely a noisy version of $RA$. Stated otherwise, $A \to RA \to RB$ is a Markov chain, hence only $n\approx l(1+\xi)h(e)$ bits are disclosed, which is not far away from the ideal $lh(e)$.

However, in the more general case where Eve gathered in $E$ some information on $A$ and $B$ by tapping the quantum channel, $A|E \to RA|E \to RB|E$ does not necessarily form a Markov chain. Instead, it must be upper bounded by the number of bits disclosed by both parties as if they were independent, $|C|=2n \approx 2l(1+\xi)h(e)$. 

Such a penalty is an effect of interactivity, as both Alice and Bob disclose some information. This can however be reduced by noticing that $RA$ and $RB$ can also be equivalently expressed by $RA$ and $R(A+B)$. The first term $RA$ gives information directly on Alice's bits $A=S_i(X_{1\dots l})$ for some slice number $i$, which are used as a part of the key. The second term $R(A+B)$ however contains mostly noise and does not contribute much to Eve's knowledge. This must however be explicitly evaluated with all the details of the QKD protocol in hands \cite{gros02:nature}.

With SEC, it is not required to use the same protocol for all slices. Non-interactive and interactive BCPs can be combined. In the particular case of slices with large $e_i$, disclosing the entire slice may cost less than interactively correcting it. Overall, the number of bits revealed is:
\begin{equation}
\label{eq:Cpractical}
|C|=\sum_i |C_i| \text{, with } |C_i|=\min \left( l, f_i(l, e_i) \right),
\end{equation}
and $f_i(l, e_i)$ the expected number of bits disclosed by the BCP assigned to slice $i$ working on $l$ bits with a bit error rate equal to $e_i$.

As $d$ grows and it becomes sufficient to only disclose the first $r$ slices so as to leave an acceptable residual error, using a practical BCP comes closer to the bound $l^{-1}|C| \geq H(K(X)|X')$. This follows from the obvious fact that $l^{-1} \sum_{i=1}^{r} |C_i| \leq r$, while the last slices can be ignored $f_i=0$, $i > r$.

%
\section{Correction of Gaussian Key Elements}
%
\label{section:construction}

We must now deal with the reconciliation of information from Gaussian variables $X \sim N(0,\Sigma)$ and $X'=X+\epsilon$, $\epsilon \sim N(0,\sigma)$. Let us first show that this problem is different from known transmission schemes, namely quantization and coded modulation. We temporarily leave out the slice estimation problem and assume that Bob wants to have most information (in Shannon's sense) about a discrete value $T(X)$, computed by Alice, given its noisy value $X'$.

In a vector quantization (VQ) system, a random input vector $X$ is transmitted over a noiseless discrete channel using the index of the closest code-vector in a given codebook. The codebook design issue has been extensively studied in the VQ literature \cite{GraNeu98}. The criterion to optimize in that case is the average distortion between $X$ and the set of reproduction vectors. In our problem, we do not have reproduction vectors since we are not interested in reproducing the continuous code but rather extracting common information.

In a coded modulation system, a binary key $k$ is sent over a continuous noisy channel using a vector $X$ belonging to a codebook in a Euclidean space. Trellis-coded modulation and lattice-based coded modulation are instances of this scheme. In this case, the information sent on the channel is chosen by Alice in a codebook, which is not true in our case.

\subsection{Design}

In this section, we present how we designed slices and slice estimators for specifically correcting Gaussian raw keys. We now assume $d=1$, that is, Alice and Bob use Gaussian key elements individually. The idea is to divide the set of real numbers into intervals and to assign slice values to each of these intervals. The slice estimators are then derived as most likelihood estimators as explained above.

For simplicity, the design of the slices was divided into two smaller independent problems. First, we cut the set of real numbers (Alice's raw key space) into a chosen number of intervals -- call this process $T(X)$. For the chosen number of intervals, we try to maximize $I(T(X);X')$. Second, we assign $m$ binary values to these intervals in such a way that slices can be corrected with as few leaked information as possible.

If the reconciliation is optimal, it produces $H(T(X))$ common bits and discloses $I_0$ bits, thus from Eq.~(\ref{eq:i0}) giving a net result of $H(T(X))-H(T(X)|X')=I(T(X);X')$ bits. Note that $S_{1\dots m}(X)$ will be an invertible function of $T(X)$. However, optimizing $I(T(X);X')$ does not depend on the bit assignment, so this is not yet relevant.

The process $T(X)$ of dividing the real numbers into $t$ intervals is defined by $t-1$ variables $\tau_1 \dots \tau_{t-1}$. The interval $a$ with $1\leq a \leq t$ is then defined by the set $\{ x \ : \ \tau_{a-1} \leq x < \tau_a \}$ where $\tau_0 = -\infty$ and $\tau_t = +\infty$. The function $I(T(X);X')$ was numerically maximized under the symmetry constrains $\tau_a=\tau_{t-a}$ to reduce the number of variables to process.

The results are displayed in Fig.~\ref{fig:t_vs_I} below. $I(T(X);X')$ is bounded from above by $\log t$ and goes to $\frac{1}{2}\log(1+\mbox{SNR})$ as $t \to \infty$.

Let us detail the expressions we evaluated. The random variable $X$ is Gaussian with variance $\Sigma^2$. $X'$ is the result of adding a random noise $\epsilon$ of variance $\sigma^2$ to $X$. Hence, the random variables $X$ and $X'$ follow the joint density function
\begin{equation*}
f_{X,X'}(x,x') = \frac{1}{2 \pi \Sigma \sigma}e^{-x^2/2\Sigma^2}e^{-(x-x')^2/2\sigma^2}.
\end{equation*}
Since $I(T(X);X')=H(T(X))+H(X')-H(T(X),X')$, we need to evaluate the following terms.
\begin{gather*}
H(T(X))=-\sum_a P_a \log P_a,\text{ with } \\
P_a=\frac{1}{2}\left( \erf \left( \frac{\tau_a}{\sqrt{2} \Sigma} \right) - \erf \left( \frac{\tau_{a-1}}{\sqrt{2} \Sigma} \right) \right), \\
H(X')=\frac{1}{2} \log 2 \pi e (\Sigma^2+\sigma^2),\text{ and } \\
H(T(X),X') = -\sum_a \int_{-\infty}^{+\infty}dx' f_a(x') \log f_a(x'),\text{ with } \\ f_a(x')=\int_{\tau_{a-1}}^{\tau_a} dx f_{X,X'}(x,x').
\end{gather*}

\begin{figure}[p]
\begin{center}\includegraphics[width=8cm]{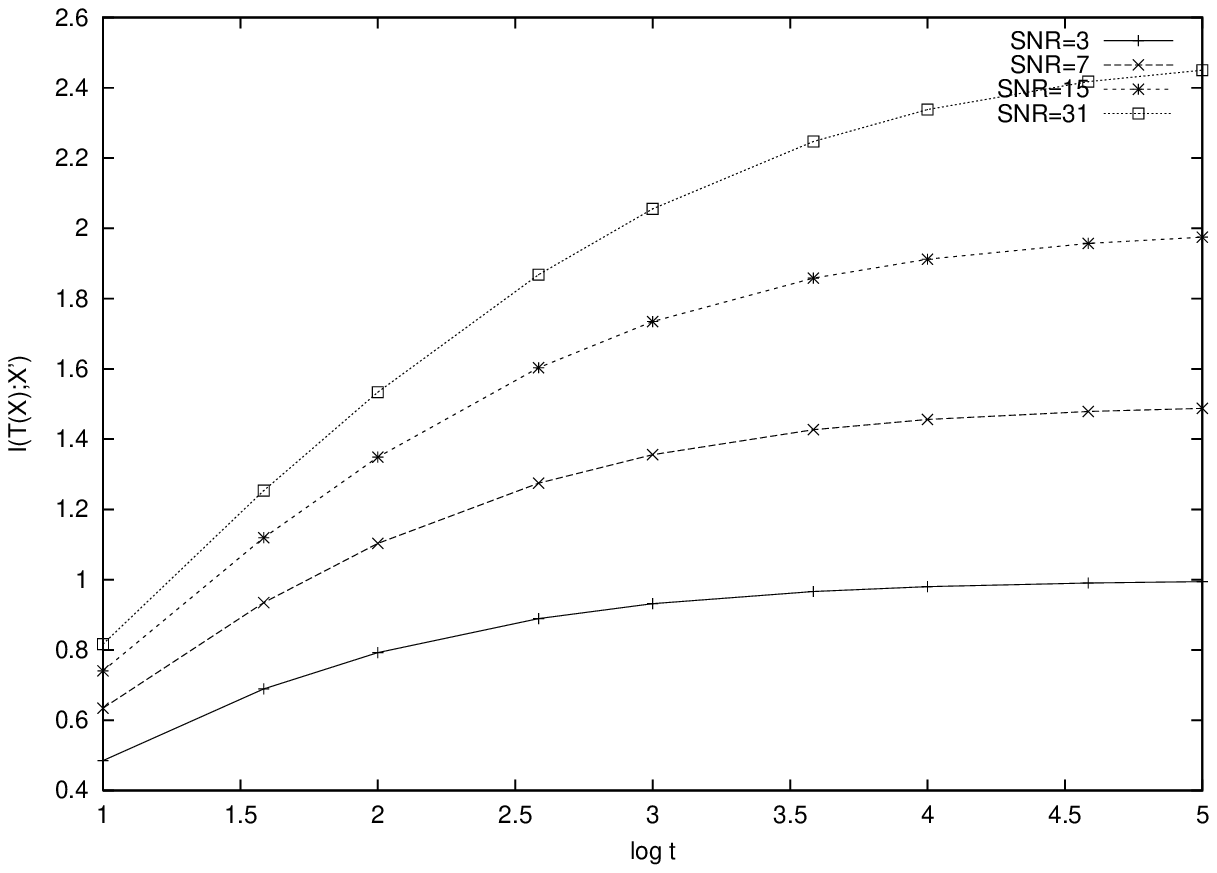} 
\end{center}
\caption{Optimized $I(T(X);X')$ as a function of $\log t$ for various signal-to-noise ratios, with $t$ the number of intervals}
\label{fig:t_vs_I}
\end{figure}

From the above procedure, we get intervals that are bounded by the thresholds $\tau_a$. The next step is to construct $m$ slices that return binary values for each of these intervals. Let us restrict ourselves to the case where $t$ is a power of two, namely $t=2^m$. We investigated several assignment methods, and it turned out that the best bit assignment method consists of assigning the least significant bit of the binary representation of $a-1$ ($0 \leq a-1 \leq 2^m-1$) to the first slice $S_1(x)$ when $\tau_{a-1} \leq x < \tau_a$. Then, each bit of $a-1$ is subsequently assigned up to the most significant bit, which is assigned to the last slice $S_m(x)$. More explicitly, 
\begin{equation}
\label{eq:les}
S_i(x) =
\left\{ \begin{array}{ll}
0 & \textrm{if } \tau_{2^i n} \leq x < \tau_{2^i n+2^{i-1}}, \\
1 & \textrm{otherwise.}
\end{array} \right.
\end{equation}
This ensures that the first slices containing noisy values help Bob narrow down his guess as quickly as possible.

\subsection{Numerical Results}

Let us now give some numerical examples in the case of a BCP optimized for a BSC, as this is the most frequent case in practice. To make the discussion independent of the chosen BCP, we evaluated $H(S_{1\dots m}(X))$ and $I_e = \sum_i h(e_i)$ for several $(m, \Sigma/\sigma)$ pairs, thus assuming a perfect BSC-BCP. (Note that, in practice, one can make use of the properties of the practical BCP chosen so as to optimize the practical net secret key rate \cite{gros02:nature}.)

\begin{figure}[p]
\begin{center}
\includegraphics[width=8cm]{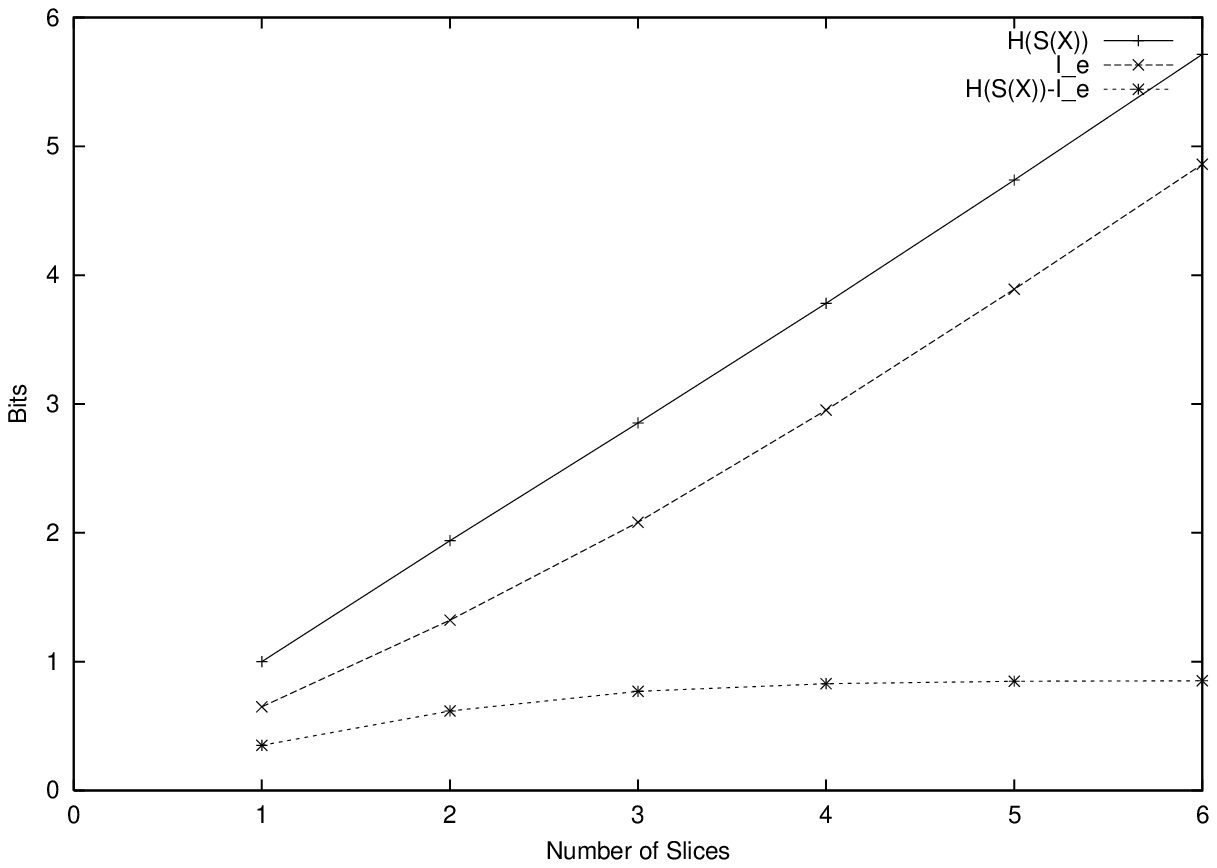} 
\end{center}
\caption{$H(S_{1\dots m}(X))$, $I_e$ and their difference as a function of the number of slices $m$ when $\Sigma^2/\sigma^2=3$}
\label{fig:snr003}
\end{figure}

Assume that the Gaussian channel has a signal-to-noise ratio of 3. According to Shannon's formula, a maximum of 1 bit can thus be transmitted over such a channel. Various values of $m$ are plotted in Fig.~\ref{fig:snr003}. First, consider the case $m=1$, that is only one bit is extracted and corrected per Gaussian value. From our construction in Eq.~(\ref{eq:les}), the slice reduces to the sign of $x$: $S_1(x)=1$ when $x \geq 0$ and $S_1(x)=0$ otherwise. Accordingly, Bob's most likelihood estimator (\ref{eq:mle}) is equivalent to Alice's slice, $\tilde{S}_1(x')=S_1(x')$. In this case, the probability that Alice's and Bob's values differ in sign is $e_1 \approx 0.167$ and hence $I_e=h(e_1) \approx 0.65$ bits. The net amount of information is thus approximately $1-0.65=0.35$ bit per raw key element.

Let us now investigate the case of $m=4$ slices, still with a signal-to-noise ratio of 3. The division of the raw key space into intervals that maximizes $I(T(X);X')$ is given in Fig.~\ref{table:tau}. Note that the generated intervals blend evenly distributed intervals and equal-width intervals. Evenly distributed intervals maximize entropy, whereas equal-width intervals best deal with additive Gaussian noise.
\begin{figure}
\begin{center}
\begin{tabular}{|ll|ll|}
\hline
$\tau_8$ & 0 & $\tau_{12}=-\tau_4$ & 1.081 \\
$\tau_9=-\tau_7$ & 0.254 & $\tau_{13}=-\tau_3$ & 1.411 \\
$\tau_{10}=-\tau_6$ & 0.514 & $\tau_{14}=-\tau_2$ & 1.808 \\
$\tau_{11}=-\tau_5$ & 0.768 & $\tau_{15}=-\tau_1$ & 2.347 \\
\hline
\end{tabular}
\end{center}
\caption{Symmetric interval boundaries that maximize $I(T(X);X')$, with $\Sigma=1$ and $\sigma=1/\sqrt{3}$}
\label{table:tau}
\end{figure}

Alice's slices follow Eq.~(\ref{eq:les}), and Bob's slice estimators are defined as usual using Eq.~(\ref{eq:mle}). The correction of the first two slices (i.e., the least two significant bits of the interval number) have an error rate that make them almost uncorrelated, namely $e_1 \approx 0.496$ and $e_2 \approx 0.468$. Then comes $e_3 \approx 0.25$ and $e_4 \approx 0.02$. Note that slice 4 gives the sign of $x$, just like the only slice when $m=1$ above. The error rate is different here because correcting slice 4 in this case benefits from the correction of the first three slices. Indeed, for $m=4$, the net amount of information is about $3.78-2.95=0.83$ bit per raw key element.

We also investigated other signal-to-noise ratios. When $\Sigma^2/\sigma^2 = 15$, Alice and Bob can share up to 2 bits per raw key element. With $m=5$, this gives a net amount of information of about $1.81$ bits per raw key element.

As one can notice, the first few error rates (e.g., $e_1$ and $e_2$) are high and then the next ones fall dramatically. The first slices are used to narrow down the search among the most likely possibilities Bob can infer, and then the last slices compose the shared secret information. Also, slices with high error rates play the role of sketching a hypothetical codebook to which Alice's value belongs. After revealing the first few slices, Bob knows that her value lies in a certain number of narrow intervals with wide spaces between them. If Alice had the possibility of choosing a codebook, she would pick up a value from a discrete list of values -- a situation similar to the one just mentioned except for the interval width. Using more slices $m>4$ would simply make these codebook-like intervals narrower.

\begin{figure}[p]
\begin{center}
\includegraphics[width=8cm]{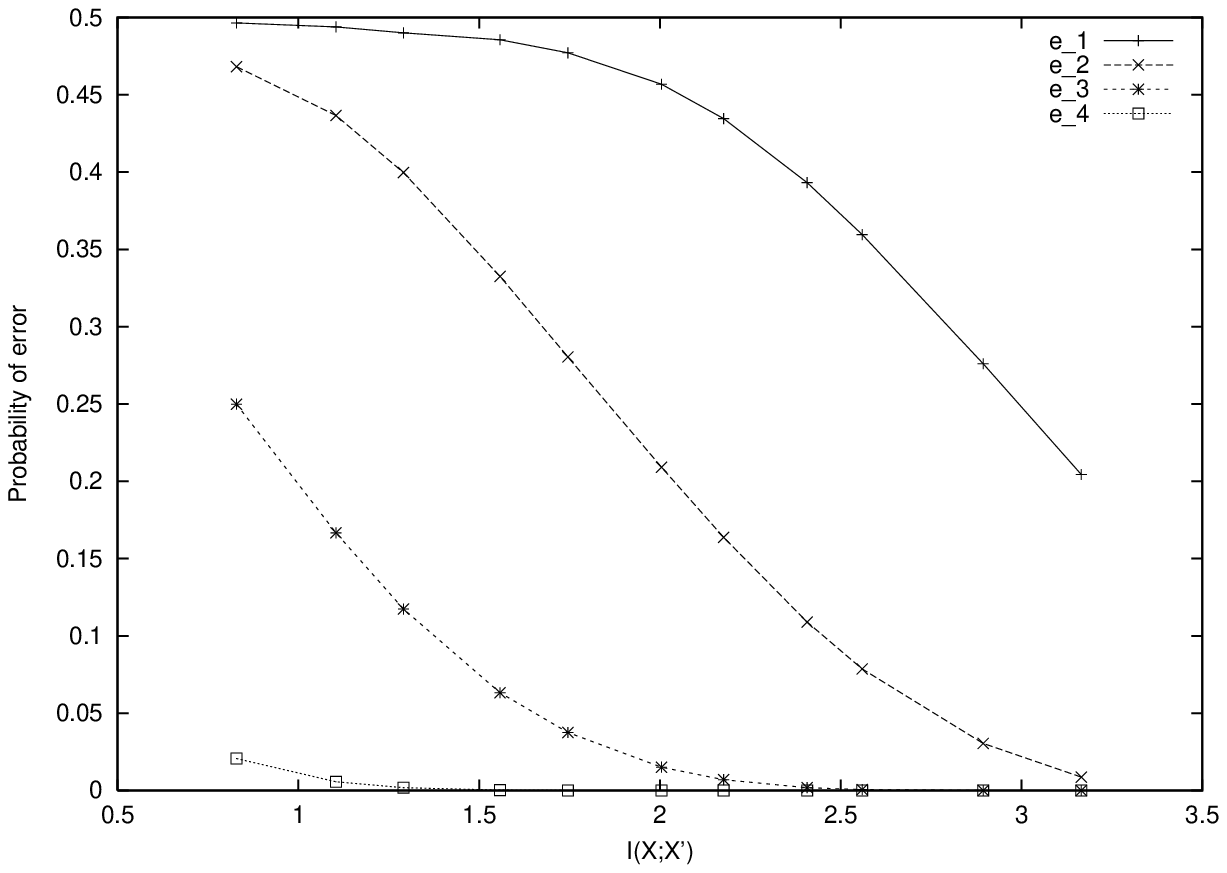} 
\end{center}
\caption{Error rates $e_{1,2,3,4}$ as a function of the channel capacity $I(X;X')$}
\label{fig:errors}
\end{figure}

In figure~\ref{fig:errors}, we show these error rates for $m=4$ when the noise level varies. From the role of sketching a codebook, slices gradually gain the role of really extracting information as their error rates decrease with the noise level.

%
\section{Conclusions}
%

Current reconciliation procedures are aimed at correcting strings of bits. A new construction for reconciliation was proposed, which can be implemented for extracting common information out of any shared variables, either discrete or continuous. This construction is then applied to the special case of Gaussian key elements, in order to complement Gaussian-modulated quantum key distribution schemes \cite{cerf00:qdgk,gros02:coherent,gros02:nature}. This might also be applied to other quantum key distribution schemes \cite{reid99,ralp00,hill00,silb00} that deal with continuous variables as well. We showed theoretical results on the optimality of our construction when applied to asymptotically large bloc sizes. Practical results about reconciliation of Gaussian key elements show that such a construction does not leak much more information than the theoretical bound. 

\bibliography{qit,rqdgk}

\appendix

%
\section{Proof of Asymptotic Optimality}
%
\label{app:asympt}

\begin{lemma}
\label{lem:list}
Let $Z=(Z_1 \dots Z_N)$ a list of $N$ random bit strings of arbitrary length, independently and uniformly distributed. The probability that a given string from the list, say $Z_j$, can be uniquely identified in $Z$ by specifying only the first $r$ bits is $(1-2^{-r})^{N-1}$.
\end{lemma}
\begin{proof}
The probability of $Z_j$ being uniquely identifiable from its first $r$ bits is the probability that no string among the $N-1$ other ones in the list starts with the same pattern. Hence, this probability is $(1-2^{-r})^{N-1}$.
\end{proof}

\begin{lemma}
\label{lem:typical}
\cite{cover} Let $X$ and $X'$ be discrete random variables distributed as $p(x,x')$ and $A^{(d)}_{\epsilon}(X,X')$ be the set of jointly typical sequences $(X^{(d)}, {X'}^{(d)})$ of length $d$. Let ${x'}^{(d)}$ be some fixed sequence in the set $A^{(d)}_{\epsilon}(X')$ of typical sequences in the marginal distribution of $X'$. Define $A^{(d)}_{\epsilon}(X|{x'}^{(d)}) = \{ x^{(d)} \ : \ (x^{(d)},{x'}^{(d)}) \in A^{(d)}_{\epsilon}(X,X') \}$. Then, $| A^{(d)}_{\epsilon}(X|{x'}^{(d)}) | \leq 2^{d(H(X^{(1)}|{X'}^{(1)})+2\epsilon)}$.
\end{lemma}

\begin{lemma}
\label{lem:asympt}
Suppose that Alice sends a discrete random sequence $X^{(d)}$ of length $d$ and Bob receives a correlated sequence ${X'}^{(d)}$, which are jointly typical $(x^{(d)},{x'}^{(d)}) \in A^{(d)}_{\epsilon}(X,X')$. Let $m = \lceil dH(X^{(1)})+\epsilon \rceil$. Let the $m$ slices $S_{1\dots m}(X^{(d)})$ be chosen randomly using a uniform distribution independently for all input values. Let $r=\lceil dH(X^{(1)}|{X'}^{(1)})+2\epsilon - \log \epsilon +1 \rceil$. Then $\forall \epsilon > 0 \ \exists D$ such that $\forall d>D$, Bob can recover $X^{(d)}$ given ${X'}^{(d)}$ and $S_{1\dots r}(X^{(d)})$ with a probability of identification failure $P_i < \epsilon$.
\end{lemma}
\begin{proof}
Alice and Bob agree on a random $S_{1\dots m}(X^{(d)})$. Assume that they draw sequences $x^{(d)}$ and ${x'}^{(d)}$ that fulfill the typicality conditions above. For the value received, Bob prepares a list of guesses: $\{ x^{(d)} \in A^{(d)}_{\epsilon}(X|{x'}^{(d)}) \}$. From Lemma~\ref{lem:typical}, this list contains no more than $N \leq 2^{dH(X^{(1)}|{X'}^{(1)})+2\epsilon}$ elements. Alice reveals $r$ slice values, with $r \geq dH(X^{(1)}|{X'}^{(1)}) + 2\epsilon-\log \epsilon +1$. From Lemma~\ref{lem:list}, the probability that Bob is unable to correctly identify the correct string is bounded as $P_i \leq 1-(1-2^{-dH(X^{(1)}|{X'}^{(1)})-2\epsilon+\log \epsilon-1})^{2^{dH(X^{(1)}|{X'}^{(1)})+2\epsilon}-1}$. This quantity goes to $1-e^{-\epsilon/2}$ when $d \to \infty$, and $1-e^{-\epsilon/2} < \epsilon/2$ for $\epsilon>0$. Therefore, $\exists D$ such that $P_i < \epsilon$ for all $d>D$.
\end{proof}

\begin{lemma}
\label{lem:discasympt}
Sliced error correction on the discrete variables $X$ and $X'$, together with an all-or-nothing BCP, leaks an amount of information that is asymptotically close to $H(X|X')$ per raw key element as $d \to \infty$, with a probability of failure that can be made as small as desired.
\end{lemma}
\begin{proof}
Using random coding arguments, lemma~\ref{lem:asympt} states that for each $d$ sufficiently large, there exists slices $S^{(d)}_{1\dots m}$ of which the first ones are to be entirely disclosed, giving $|C| \leq l^{(d)}(dH(X^{(1)}|{X'}^{(1)}) + 2\epsilon-\log \epsilon +2$. The number $l^{(d)}$ of key elements of dimension $d$ is $l^{(d)}=l^{(1)}/d$ with $l^{(1)}$ the number of raw key elements. Hence $|C| \leq l^{(1)}(H(X^{(1)}|{X'}^{(1)}) + d^{-1}(2\epsilon-\log \epsilon +2))$. Regarding the probability of failure, there are two sources of possible failure: the failure of identification $P_i$ and the fact that $(x^{(d)},{x'}^{(d)}) \notin A^{(d)}_{\epsilon}(X,X')$. From Lemma~\ref{lem:asympt} and from the AEP, both probabilities are upper bounded by $\epsilon$. Therefore, the total failure probability behaves as $O(\epsilon)$ when $\epsilon \to 0$.
\end{proof}

\begin{theorem}
\label{th:asympt}
Sliced error correction on the random variables $X$ and $X'$, together with an all-or-nothing BCP, can make $H(K(X))-l^{-1}|C|$ as close as desired to $I(X;X')$.
\end{theorem}
\begin{proof}
If $X$ is discrete, let $K(X)=X$, otherwise set $K(X)=\hat{X}$, with $\hat{X}$ a quantized approximation of $X$. Similarly, let $\hat{X}'=X'$ when $X'$ is discrete or approximate it with a discrete variable $\hat{X}'$ otherwise. For any $\hat{\epsilon}>0$, there exits $\hat{X}$, $\hat{X}'$ such that $I(\hat{X};\hat{X}') \geq I(X;X')-\hat{\epsilon}$ \cite{cover}.

By applying Lemma~\ref{lem:discasympt} on $\hat{X}$ and $\hat{X}'$, we have
$|C| \leq l(H(\hat{X}|\hat{X}')+\epsilon')$ for any $\epsilon'>0$. Therefore,
\begin{equation}
\begin{split}
H(K(X))-l^{-1}|C| &\geq H(\hat{X}) - H(\hat{X}|\hat{X}') - \epsilon' \\
&\geq I(\hat{X};\hat{X}')-\epsilon' \\
&\geq I(X;X')-\epsilon'-\hat{\epsilon}.
\end{split}
\end{equation}
\end{proof}

\begin{corollary}
\label{cor:practical}
If we use a practical BCP instead of disclosing the slice bits whenever this would leak less than $l$ bits, the conclusion of Th.~\ref{th:asympt} still applies.
\end{corollary}
\begin{proof}
Assume that we can predict how many bits the practical BCP discloses, for instance given an estimate of the bit error rate. Disclosing a slice entirely, as done in Lemma~\ref{lem:asympt}, reveals $l$ bits. Whenever the practical BCP is expected to disclose less than $l$ bits (e.g., when the bit error rate is low), we can use it instead of disclosing the entire key without increasing $|C|$.
\end{proof}


\end{document}